# GenAITEd Ghana: A First-of-Its-Kind Context-Aware and Curriculum-Aligned Conversational AI Agent for Teacher Education

Matthew Nyaaba[1], Patrick Kyeremeh[2], Macharious Nabang[3], Bismark Nyaaba Akanzire[4], Sakina Acquah[5], Cyril Ababio Titty[6], Kotor Asare[7], Jerry Etornam Kudaya[8]

[1] Department of Educational Theory and Practice, University of Georgia, Athens, GA, USA
[2] St. Joseph's College of Education, Bechem, Ghana
[3] Bagabaga College of Education, Tamale, Ghana
[4] Gambaga College of Education, Gambaga, Ghana
[5] University of Education, Winneba, Ghana
[6] Komenda College of Education, Komenda, Ghana
[7] University of Skills Training and Entrepreneurial Development, Kumasi, Ghana
[8] Generative AI for Education and Research in Africa (GenAI-ERA)

**Abstract**

Global frameworks increasingly call for Responsible Artificial Intelligence (AI) in education, yet they provide limited guidance on how ethical, culturally responsive, and curriculum-aligned AI can be operationalized within functioning teacher education systems, particularly in the Global South. This study addresses this gap through the design and evaluation of *GenAITEd Ghana*, a context-aware, region-specific conversational AI prototype developed to support teacher education in Ghana. Adopting a Design Science Research approach, the study developed GenAITEd Ghana as a school-mimetic digital infrastructure aligned with the organizational logic of Ghanaian Colleges of Education. The platform provides NaCCA-aligned course environments based on users' institutional affiliation, academic year, semester, and course specialization. Teacher educators create course-specific AI agents and invite student teachers into individual or collaborative learning spaces using cryptographic passkeys. The system operates as a multi-agent, retrieval-augmented conversational AI that coordinates multiple AI models for curriculum-grounded dialogue, automatic speech recognition, voice synthesis and cloning, and multimedia processing. Two complementary prompt pathways were embedded: system-level prompts enforcing curriculum boundaries, ethical constraints, retrieval scope, and teacher-in-the-loop oversight, and interaction-level semi-automated prompts that structure live pedagogical dialogue through clarification, confirmation, and guided response generation. Evaluation findings show that the system features and prompt logics addressed key Responsible AI framework requirements, including transparency, accountability, cultural responsiveness, privacy, and human oversight. Human expert evaluations further indicated that *GenAITEd Ghana* is pedagogically appropriate for Ghanaian teacher education and consistently perceived the system as capable of promoting student engagement while preserving educators' professional authority. However, some implementation challenges were noted, particularly regarding the successful deployment of teacher voice cloning and avatar generation for AI agents. Experts also highlighted the risk of student teachers over-relying on AI agents without sufficiently engaging other domains of learning that require human judgment, social interaction, and affective engagement. The study therefore advocates for the scalable advancement of context-aware educational AI through enhanced model integration, sustained professional development, and critical AI literacy for both teachers and student teachers.

**Introduction**

Artificial Intelligence (AI) has emerged as a defining force in education, opening new frontiers for personalization, adaptive assessment, and instructional automation. Yet, the overwhelming dominance of Western-trained language models has created an epistemic asymmetry in educational AI development. Most large-scale systems are trained on English-language, Euro-American datasets, resulting in contextual misalignment when deployed in African learning environments (Eke et al., 2023; Kiemde & Kora, 2022). Consequently, these systems often fail to capture the linguistic diversity, pedagogical culture, and curricular frameworks of the Global South. Recognizing these challenges, leading international organizations have called for ethical, inclusive, and culturally grounded AI development. The OECD Principles on AI emphasize the need for human-centered and trustworthy systems that respect democratic values and cultural diversity (OECD, 2019). Likewise, UNESCO's *Recommendation on the Ethics of Artificial Intelligence* (2021) and its *Guidance for Generative AI in Education and Research* (2023) advocate for equitable participation in AI creation, with explicit attention to linguistic inclusion and local curriculum integration. These frameworks collectively underscore that the ethical future of AI in education depends on balancing technological innovation with contextual and cultural responsiveness.

Globally, countries are localizing AI ecosystems to align with national education priorities. Examples such as Squirrel AI in China, ViLLE in Finland, and Smart Mentor in Japan illustrate a growing shift toward context-aware, curriculum-linked AI platforms (Naseri & Abdullah, 2025). Similarly, Estonia's AI Leap 2025 and India's YUVAi initiatives integrate AI literacy and responsible design into educational reforms (Education Estonia, 2025; Gichuki, 2025). In parallel, OpenAI's global democratization initiative (OpenAI, 2025) encourages country-specific model integration to empower localized innovation, highlighting the need for AI systems that reflect the social, cultural, and linguistic realities of each region. This global push is leading to the creation of Large Language Models (LLMs) and specialized AI tools tailored to specific national requirements. In Asia, Japan's Fugaku-LLM leverages its domestic supercomputing power to enhance capabilities in the Japanese language, specifically to avoid reliance on foreign technology (Tokyo Institute of Technology et al., 2024). Similarly, South Korea's HyperCLOVA X Think is being developed to excel in inference and reasoning within the Korean linguistic and societal context (Do-yun, 2025). The need for linguistic and cultural specificity is also paramount in India, where the Airavata project is instruction-tuned for Hindi and other Indic languages (Gala et al., 2024), and in Russia, where GigaChat serves as a multimodal chatbot focusing on superior Russian-language performance (GigaChat, 2025).

Within this global movement, Ghana has positioned itself as an African leader. In 2025, the Ministry of Education, in partnership with GES, NaCCA, NTC, NaSIA, and CENDLOS, launched subject-specific AI applications for Senior High Schools (SHS) through PlayLab AI, supporting over 68,000 teachers and 1.4 million students (Gichuki, 2025). However, as with the earlier *21st Century Teacher Educator GPT* (Nyaaba, 2025), these tools rely on external commercial platforms, raising issues of data sovereignty, sustainability, and infrastructural dependence. In response to these gaps, the present study introduces GenAITEd Ghana, a blueprint prototype of a context-aware and region-specific conversational AI system designed specifically for teacher education. Grounded in Ghana's NaCCA curriculum, institutional structures, and linguistic realities, the prototype demonstrates the functional design and operational feasibility of a locally governed educational AI architecture.

The study foregrounds the system's core functionalities, including curriculum activation, teacher-customized AI agents, structured conversational prompting, and teacher-in-the-loop oversight, and evaluates these features through alignment with global Responsible AI frameworks and expert human review. In doing so, the study contributes a foundational design artifact that advances discussions on AI sovereignty, contextual relevance, and responsible educational AI development in the Global South. An earlier version of this work was accepted for presentation at the 3rd UNESCO Global Forum on the Ethics of Artificial Intelligence (Bangkok, Thailand, June 24–27, 2025), reflecting its relevance to ongoing global discourse on responsible, inclusive, and locally governed AI ecosystems. However, this present study extends this earlier version by providing a detailed account of the system's architecture, prompt-engineering strategy, and evaluation procedures. Accordingly, the study is guided by the following research questions:

1. How does the GenAITEd Ghana prototype demonstrate functional integration and operational feasibility across its core components?
2. To what extent does the GenAITEd Ghana prototype align with international and continental frameworks for Responsible AI in Education?
3. How do teacher-education experts perceive the pedagogical alignment of the GenAITEd Ghana prototype?

## 2. Literature Review

### 2.1 Global Policy Frameworks for Responsible AI in Education

International organizations such as the Organisation for Economic Co-operation and Development (OECD), United Nations Educational, Scientific and Cultural Organization (UNESCO), and the African Union (AU) have developed frameworks to guide the ethical and equitable use of artificial intelligence (AI) in education. The OECD's *AI Principles* (OECD, 2019) emphasize that AI systems should be *innovative, trustworthy, transparent,* and *inclusive* to ensure that technological advancement contributes to human-centered development. Its 2023 report, *Opportunities, Guidelines and Guardrails on Effective and Equitable Use of AI in Education*, calls for governments to strengthen policy ecosystems that promote equity, teacher capacity building, and ethical accountability (OECD, 2023).

Similarly, UNESCO (2021) issued *AI and Education: Guidance for Policy-Makers*, advocating for a human-centered approach that prioritizes cultural context, teacher agency, and data ethics in the digital transformation of education. The 2023 *Guidance for Generative AI in Education and Research* expanded this framework to include emerging generative models, stressing that AI should empower educators rather than replace them (UNESCO, 2023). In Africa, UNESCO's *Artificial Intelligence Needs Assessment Survey* (2021) and the AU's *Continental Artificial Intelligence Strategy* (African Union, 2024) both emphasize localized data governance, inclusivity, and education-led ethical AI development. These frameworks affirm that responsible AI integration must be guided by *cultural responsiveness*, *ethical governance*, and *local ownership*. However, most remain conceptual, lacking operational mechanisms for implementation in low-resource or multilingual contexts such as Ghana.

### 2.2 Country-specific Sovereign AI tools

The global landscape of large language models (LLMs) is rapidly evolving beyond the initial dominance of Western systems, with numerous nations prioritizing the development of sovereign AI tailored to local languages and cultural contexts. South Korea, for example, is advancing its capabilities with the launch of HyperCLOVA X Think by Naver, a model specializing in inference and reasoning with a noted strength in the nuances of the Korean language (Do-yun, 2025). Similarly, Russia's Sberbank introduced GigaChat, a multimodal chatbot positioned as a homegrown alternative to Western models, emphasizing its superior communication and processing skills in Russian ("GigaChat," 2025). East Asia is also contributing significantly, with Japan leveraging its high-performance computing infrastructure to create the Fugaku-LLM, which is specifically enhanced for the Japanese language (Tokyo Institute of Technology et al., 2024). Meanwhile, the AI4Bharat initiative in India is tackling the complexity of the subcontinent's linguistic diversity with Airavata, an instruction-tuned LLM focused on Hindi and other Indic languages (Gala et al., 2024). This collective focus highlights a worldwide effort to create AI tools that are not only technologically advanced but also linguistically and culturally accurate for domestic use.

AI innovation across Africa has primarily focused on increasing access to learning resources amid high student–teacher ratios and infrastructural constraints. Tools such as SuaCode, AutoGrad, and Kwame for Science exemplify early advances in AI-supported education (Boateng et al., 2024). *Kwame for Science*, a bilingual AI teaching assistant, extended domain-specific tutoring to science education, combining curated knowledge sources with national examination datasets. Over eight months, it served 750 users across 32 countries (15 in Africa) with 1.5 K student questions, achieving 87.2% top-three accuracy (Boateng et al., 2024). These efforts demonstrated that AI can democratize access to quality education, particularly when localized and bilingual design principles are applied. However, most of these tools are student-facing, designed to deliver content or provide question-answering support, rather than empowering teachers as reflective practitioners. Few AI systems have been created to facilitate *teacher education, professional development,* or *curriculum-aligned instructional design* in African contexts. As a result, existing AI applications often fail to address pedagogical depth, culturally grounded dialogue, or the role of teacher agency in AI-mediated environments.

**MagicSchool.ai**

Across empirical studies focused on MagicSchool.ai, teachers and preservice teachers report accelerated lesson planning, expanded idea generation, and practical support for differentiation and assessment authoring, accompanied by a persistent need for prompt specificity, verification, and contextualization. Content analyses of AI-generated lessons indicate minimal alignment with evidence-based UDL/UDT principles, particularly for engagement, representation, and expression, underscoring risks of generic objectives, shallow cognitive demand, and superficial differentiation without targeted edits (Lammert et al., 2024). In preservice assessment design, MagicSchool use improved time efficiency and facilitated varied item types and rubric creation, yet participants emphasized the importance of checking validity, cultural fit and avoiding overreliance, with prompt engineering emerging as a moderating skill (Coşgun, 2025). In-service EFL teachers similarly valued workload reduction and differentiated materials but stressed iterative refinement to achieve pedagogical soundness and school-specific standards alignment (Setyaningsih et al., 2024; ElSayary et al., 2025). Participatory co-design work and primary-level applications further illustrate human–AI co-agency and the production of deployable artifacts when iterative feedback loops are present, though formal artifact quality ratings and classroom enactment metrics are sparse (Tilak et al., 2024; Yudono & Widya, 2025). Discipline-specific explorations (e.g., math personalization) point to affordances for readability control and contextual reframing, while documenting limitations that mirror the broader pattern: AI drafts as decision-support that require teacher expertise to achieve rigorous instructional design (Beauchamp et al., 2025).

Supplementary evidence from comparable AI lesson-planning systems strengthens the pedagogical and teacher-education implications relevant to MagicSchool.ai. Large-scale, human-in-the-loop deployments (e.g., Shiksha copilot) demonstrate substantial reductions in planning time and stress and a reported shift toward activity-based pedagogy, with curator-mediated curation/customization workflows helping teachers adapt AI outputs to local contexts (Dennison et al., 2025). Experimental and lesson-study designs show pre/post improvements in instructional design artifacts and growth in technological pedagogical content knowledge (TPACK) when GenAI is scaffolded within collaborative planning and clear usage guidelines, though the benefits can wane without structured facilitation in later iterative phases (Chen et al., 2025; Cheng, 2025; Sun & Huang, 2025). Artifact rating studies in music education reveal that while teachers struggle to distinguish AI from human lesson plans, human-authored plans still score higher on quality, and raters rely on specificity and classroom knowledge cues, evidence that quality assurance rubrics and blind expert ratings are critical for teacher education using AI (Cooper, 2024). Taken together, these findings suggest that MagicSchool.ai is pedagogically useful for drafting and efficiency, but demonstrable gains in teacher knowledge/skills (e.g., assessment literacy, differentiation quality, pedagogical reasoning) depend on structured PD that embeds backward-design/UDL checklists, artifact-rating protocols, cognitive reasoning scaffolds, and observation rubrics tied to AI-generated plans to ensure validity, coherence, and transfer to classroom practice (Lammert et al., 2024; Coşgun, 2025; Setyaningsih et al., 2024).

## 2.3 Culturally Responsive and Linguistically Inclusive AI Design

Recent scholarships have begun to emphasize culturally responsive and linguistically inclusive AI as a foundation for equitable educational innovation. *Nyaaba and Zhai (2025)* introduced the *Culturally Responsive Lesson Planner (CRLP) GPT*, a generative AI tool based on Culturally Responsive Pedagogy (CRP) that co-constructs lesson plans through interactive prompting. Expert review revealed that CRLP outputs exhibited greater cultural relevance, factual accuracy, and curriculum alignment than those produced by generic GPT models. Extending this approach, *Nyaaba (2025)* developed the *21st Century Teacher Educator GPT* using a Glocalized Generative AI framework that fuses Ghana's National Teacher Education Curriculum Framework (NTECF) with UNESCO's AI ethics principles. This system supports pre-service teachers through bilingual interfaces in English, Twi, Dagbani, Mampruli, and Dagaare, demonstrating that localizing language, curriculum, and pedagogy can significantly improve educational AI relevance and equity.

Language inclusion has also been explored in *Glover-Tay and Korsah (2024)*, who developed a Twi-interfaced ChatGPT to support self-learning among rural junior high students. Although performance was constrained by limitations in automatic speech recognition, the study found strong user acceptance for local-language interaction. Similarly, *Garba et al. (2024)* developed a transformer-based text generation model for Nigerian Pidgin, achieving BLEU scores of 0.56 and perplexity scores of 43.26, indicating the viability of extending generative AI to under-resourced languages. In the realm of ethics, *Eke et al., (2023)* and *Kiemde and Kora (2022)* argued that education is central to responsible AI development in Africa. They emphasized that African moral philosophies, communal values, and humanistic ethics should shape AI governance rather than relying solely on imported Western frameworks. These perspectives reinforce the notion that AI design must move beyond translation or adaptation to truly embody the linguistic and cultural worldviews of African societies.

## 2.4 Identified Gaps and Research Direction

Despite progress in global frameworks, country-specific AI models and local innovations, three interrelated gaps persist in AI applications for African education. Few AI systems integrate national curriculum standards, teacher education frameworks, or context-specific pedagogical realities into their core design. Most AI tools omit indigenous philosophies and cultural markers that shape instructional practice in African classrooms. Linguistic Gap. Limited inclusion of African languages, particularly in voice-based and interactive systems, constrains equitable access and reduces authenticity. Consequently, there is an urgent need for a teacher-centered, context-aware, culturally grounded, and linguistically inclusive AI ecosystem that empowers educators to co-design, interact with, and ethically govern AI tools. The GenAITEd Ghana blueprint responds to this gap by integrating Ghanaian curricula, multilingual voice interaction, and culturally responsive pedagogy into an interactive AI environment for teacher education. This study therefore contributes to the growing discourse on *Responsible AI for Africa*, operationalizing the principles advanced by the OECD, UNESCO, and AU within a localized technological and pedagogical framework.

## 4 Method

This study adopted a design science research approach (Peffers et al., 2007; Nyaaba & Zhai, 2025) to guide the development and evaluation of GenAITEd Ghana, a context-aware conversational AI system for teacher education. Design science was appropriate because the study focused on the design, testing, and evaluation of a technological artifact, rather than on measuring instructional outcomes. The research followed a three-phase procedure. *Phase 1* focused on system design and prompt development. *Phase 2* examined functional feasibility through simulated system use. *Phase 3* evaluated the prototype through alignment with global Responsible AI frameworks and expert review. Phase 2 addressed Research Question 1, while Phase 3 addressed Research Questions 2 and 3. Detailed descriptions of the system architecture, prompt mechanisms, and functionalities are presented in the subsequent sections.

## 4.1 Procedure (Phases)

**Phase 1: System Design and Prompt Development**

In Phase 1, the study focused on the overall system architecture and initial development of GenAITEd Ghana. The platform was designed as a web-based, multi-agent conversational AI system that integrates a backend for AI orchestration and curriculum retrieval with a frontend for teacher and student interaction. The architecture combines retrieval-augmented generation (RAG) with large language models to ensure that AI responses are grounded in curriculum-approved and locally relevant content. Within this architecture, prompt design was embedded as a core system component. Two types of prompts were developed at this stage. System-level prompts were incorporated into the backend to define agent roles, curriculum boundaries, retrieval rules, and governance constraints. Interaction-level, semi-automated prompts were designed to structure live conversations by guiding clarification, confirmation, and response generation during user interaction. These prompts operate within the conversational layer of the system and shape how AI agents respond to users. This phase established the foundational architecture, prompt structure, and system logic required for subsequent functional testing and evaluation.

**Phase 2: Functional Feasibility Testing**

In the second phase, the system and the embedded prompts were tested to examine whether they functioned as intended. Testing involved logging into the system as different user roles (teachers and students), activating curriculum agents based on college, academic year, and semester, creating AI agents and learning groups, and engaging in text- and voice-based conversations. This phase examined how the prompt texts operated during real interaction, including how they guided AI responses, structured conversations, and supported curriculum alignment. Phase 2 primarily addressed Research Question 1, which focused on system integration and operational feasibility.

**Phase 3: Evaluation**

In the third phase, the system and its prompt mechanisms were evaluated using two approaches. First, the system was examined against global Responsible AI frameworks and compared with similar AI education tools to assess governance, transparency, and contextual relevance. Second, human expert evaluation was conducted with teacher education experts in Ghana, who interacted with the system and reviewed its functionality and relevance. This phase assessed how the prompt design supported responsible AI use and pedagogical alignment, addressing Research Questions 2 and 3.

## 4.2 System Overview

The project was implemented as a full-stack web application designed to demonstrate feasibility, modularity, and localization potential.
It comprised two tightly coupled components:

1. Backend Server (Node.js/Express) -manages API routing, database logic, AI orchestration, and multilingual voice processing.

2. Frontend Client Interface (JavaScript/HTML/CSS) - provides teachers with real-time text and voice interaction, curriculum browsing, and analytics visualization.

Communication between client and server followed a *REST + WebSocket* hybrid architecture, supporting both synchronous and streaming interactions. The system was designed for open scalability, allowing replication in other regions or adaptation for additional African languages.

**4.3 Data Sources**

*4.3.1 Curriculum and Educational Data*

Pedagogical grounding was achieved using Ghana's NaCCA-approved curriculum for Early Grade, Upper Primary, and Junior High School (JHS) levels. Official syllabi and competency frameworks were parsed and standardized into JSON metadata with the following fields: *subject | grade_level | strand | sub_strand | core_competency | learning_indicator.* This structure enabled dynamic prompt contextualization so that every AI-generated output, lesson outline, assessment item, or reflection, remained aligned with Ghana's competency-based curriculum. The curriculumService.js module indexed this dataset to enable semantic search and retrieval.

*4.3.2 Voice and Linguistic Resources*

The voiceService.js component integrated open-source automatic speech recognition (ASR) and text-to-speech (TTS) models fine-tuned for English and Ghanaian languages (*Twi*, *Dagbani*, *Ewe*). Voice data consisted solely of synthetic or anonymized system test recordings, ensuring that no human-subject voice data were retained. These resources were used only to evaluate latency, word-error rate (WER), and multilingual intelligibility. This multilingual design demonstrates how localized AI can foster accessibility and inclusivity for educators in diverse linguistic regions.

**4.4 Architecture**

*4.4.1 Backend Architecture*

The backend employed Node.js (v18) and Express.js for modular scalability.
The main server file, server.js, initializes middleware for CORS, JSON parsing, and authentication before mounting the API routes (see Table 1).

Core directories and functions:

- routes/ – defines endpoints (teacherRoutes.js, chatbotRoutes.js, adminRoutes.js).
- services/ – houses reusable logic (aiService.js for AI calls; voiceService.js for audio processing).
- database/ – implements a *PostgreSQL* relational schema ensuring referential integrity.

**Table 1**

*Primary database tables*

| Table | Function |
| --- | --- |
| users | Stores role metadata for teachers and administrators. |
| conversations | Logs AI session data (text + audio) with timestamps. |
| curriculum_items | Indexed NaCCA curriculum records for retrieval. |
| analytics | Tracks latency, token counts, and system performance. |

This modular backend architecture allows horizontal scaling and facilitates deployment in low-bandwidth environments using containerized micro-services.

*4.4.1.1 Routing and API Management.* The /routes directory provides RESTful and streaming endpoints connecting user interactions to backend services (see Table 2). The modular routing structure enhances code readability, access control, and scalability. This routing design ensures clear separation of concerns and efficient middleware validation, enabling secure communication between frontend modules, AI services, and curriculum databases.

**Table 2**

*Routing and API Management*

| File | Description |
| --- | --- |
| *adminRoutes.js* | Handles administrative operations including curriculum uploads, system logs, and configuration management. |
| *teacherRoutes.js* | Manages teacher accounts, authentication, and data retrieval for custom GPT creation. |
| *userRoutes.js* | Governs user-level operations such as registration and profile management. |
| *gptRoutes.js* | Interfaces with the AI service layer to handle all generative model requests and prompt exchanges. |
| *groupRoutes.js* | Supports passkey-based group creation and shared GPT collaboration sessions. |
| *fileRoutes.js* | Manages file uploads, curriculum import/export, and multimedia resources. |
| *imageRoutes.js* | Integrates AI image-generation endpoints for visual learning materials. |
| *liveChat.js* | Establishes WebSocket channels for real-time conversational AI sessions, including voice streaming. |

*4.4.2 AI Service Layer*

At the heart of the GenAITEd Ghana platform lies the aiService.js module, which functions as the central intelligence hub responsible for orchestrating interactions across multiple large language model (LLM) providers. While OpenAI's GPT-4 Turbo serves as the primary inference engine, the system is also equipped to dynamically route requests to alternative endpoints such as *Anthropic Claude, Groq, or Hugging Face Transformer*s, depending on operational factors such as context length, cost efficiency, and latency requirements. The inference process unfolds through a structured pipeline designed to ensure pedagogical coherence and cultural fidelity. First, the system parses the user's query, identifying its instructional intent and linguistic nuances. Next, it retrieves the relevant segment of Ghana's national curriculum, matching the query to the appropriate subject, grade level, competency, and learning indicator.

This data is then embedded within a tailored prompt template that fuses educational logic with cultural specificity, integrating familiar Ghanaian names, scenarios, proverbs, and classroom terminologies. Once the prompt is fully constructed, it is dispatched to the most suitable LLM endpoint for processing. The AI response, whether in structured JSON format, natural text, or synthesized audio, is then returned to the user interface, maintaining contextual clarity and instructional relevance. This intelligent orchestration ensures that every AI-generated output is not only technically robust but also culturally grounded, pedagogically aligned, and responsive to the realities of teaching and learning in Ghanaian and West African classrooms. It reflects a deliberate move away from one-size-fits-all AI applications toward systems that honor local knowledge, linguistic diversity, and curriculum sovereignty.

*4.4.2.1 Prompt Engineering Module.* The /prompts directory defines the *instructional logic* that guides all AI interactions. Each file modularizes specific aspects of contextual prompting, ensuring that responses remain culturally, linguistically, and pedagogically grounded. Following this, the development of GenAITEd Ghana is underpinned by a two-layer prompt engineering framework. The first layer consists of system-level prompts that engineer the multi-agent retrieval-augmented generation (RAG) backend of the AI system. These system prompts define agent roles, curriculum constraints, ethical guardrails, institutional rules, and inter-agent communication protocols, ensuring that all AI outputs remain aligned with Ghana's NaCCA curriculum and teacher-in-the-loop governance structures. This layer operates persistently in the background and shapes how AI agents retrieve, reason over, and generate responses across the platform (See Table 3).

**Table 3**

*Prompt Engineering*

| File | Function |
| --- | --- |
| masterPrompt.js | Establishes the system's global instruction set, defining teacher–AI interaction boundaries, ethical constraints, and tone of communication. |
| contextModifiers.js | Dynamically adjusts prompts using local variables such as grade level, subject, or Ghanaian cultural context (e.g., Twi proverbs, classroom analogies). |
| responseFormatter.js | Structures AI outputs for clarity and voice-readability, formatting responses into instructional steps or summaries. |
| voiceInstructions.js | Embeds voice interaction parameters (pausing, emphasis, multilingual cues) for text-to-speech processing. |
| webSearchInstructions.js | Governs API calls and structured retrieval for context expansion, ensuring responsible integration of external knowledge. |
| index.js | Orchestrates prompt compilation and dispatching to aiService.js. |

The second layer comprises **interaction-level, semi-automated conversational prompts** that dynamically structure user–AI conversations during live interactions. Drawing on the interactive semi-automated prompting approach proposed by **Nyaaba and Zhai (2025)**, this layer reformulates user inputs into pedagogically coherent conversational prompts that reflect course context, semester progression, and instructional intent. Unlike static prompt templates, these semi-automated prompts adapt in real time, supporting natural dialogue while reducing ambiguity and minimizing AI hallucinations. Figure 1 shows the interactive semi-automated conversational prompting process between a student and a teacher AI agent. The student initiates interaction by typing or speaking, after which the AI agent asks curriculum-based questions, generates follow-up prompts, and summarizes the student's inputs for confirmation. Once the student confirms or revises the summary, the AI produces contextually relevant responses aligned with the course content. This structured dialogue supports accurate interpretation, reduces ambiguity, and enables real-time, transcript-based interaction.

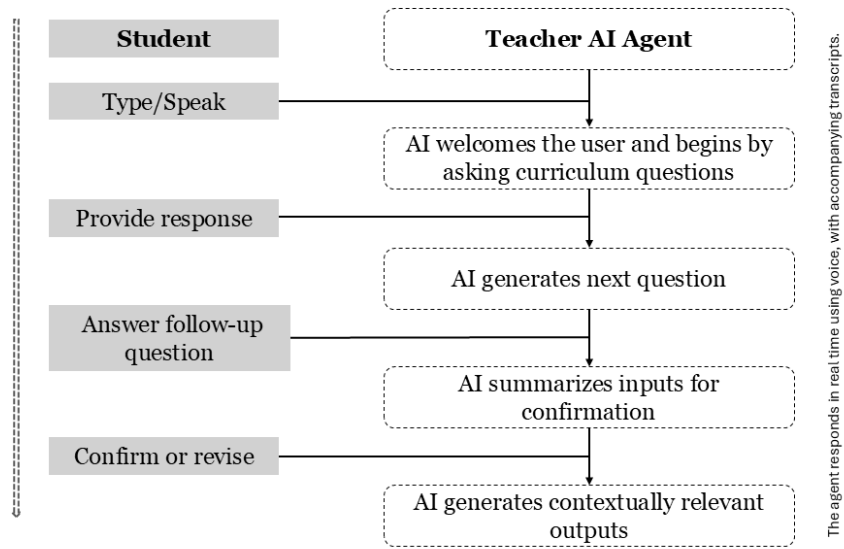

**Figure 1:**
*Interactive Semi-Automated Conversational Prompting Workflow (Adopted from Nyaaba and Zhai, 2025)*

*4.4.3 Voice Service Layer*

Bidirectional voice streaming was realized through *WebSocket channels* to support real-time interaction. The module integrates:

- ASR: OpenAI Whisper v3 for transcription.
- TTS: ElevenLabs API and local FastSpeech 2 engine for audio generation.
- Session Control: LiveKit library for managing multi-user voice rooms.

Average end-to-end response latency was ≈ 1.8 s, meeting conversational thresholds even on 3G networks. All latency and accuracy statistics were system-level diagnostics, not human performance data.

*4.4.4 Curriculum Service Layer*

At the heart of GenAITEd Ghana's contextual intelligence is the curriculumService.js module, which serves as a vital bridge between Ghana's official curriculum and the AI engine. This component ensures that every AI response is not just generative but grounded in the pedagogical expectations of the NaCCA. It achieves this by enabling keyword searches and semantic matching across learning indicators, retrieving exemplar activities and assessment templates, and exporting outputs in structured JSON formats for transparency and analytical tracking. In practical terms, this means the AI doesn't generate content in a vacuum, it responds with awareness of subject strands, learning outcomes, and competency expectations. By keeping all outputs aligned with the national curriculum, the system upholds its commitment to responsible, localized, and educationally valid AI for Ghanaian classrooms.

**4.5 Frontend Design**

The GenAITEd Ghana platform was carefully designed with simplicity and efficiency in mind. Its frontend interface built using vanilla JavaScript, HTML5, and CSS3, ensures that even devices with limited processing power can access the full range of features. At its core, the platform offers a user-friendly chat interface that supports both voice and text interaction, a curriculum browser for navigating Ghana's NaCCA standards and generating AI-assisted lesson plans, and an analytics dashboard that helps visualize key metrics such as response time and system usage. All activities are handled smoothly in the background through asynchronous communication with the backend, allowing for real-time updates and system tracking. Importantly, the interface is fully responsive, adapting easily to mobile devices, and includes offline functionality through service-worker caching, making it ideal for rural or bandwidth-constrained environments common across West Africa. This design reflects a deep commitment to educational access, technological resilience, and cultural practicality.

*4.5.1 GenAITEd Ghana System Architecture and Pedagogical Workflow*

Figure 2 summarizes and presents the operational architecture of GenAITEd Ghana, illustrating how the system was intentionally designed to mirror the organizational and pedagogical structure of Ghanaian Colleges of Education while embedding Responsible AI principles at both system and interaction levels. The figure illustrates a scalable model of context-aware conversational AI that integrates curriculum governance, collaborative learning, and ethical safeguards into a coherent educational ecosystem. The architecture foregrounds teachers as instructional authorities and positions artificial intelligence as a mediated pedagogical partner rather than an autonomous instructional agent. At the access level, both teachers and student teachers log in through institution-based authentication, reinforcing institutional governance and curriculum sovereignty. Teachers enter the system as course leaders and create subject- and specialization-specific learning spaces aligned with Ghana's teacher education tracks, including Early Grade, Upper Grade, JHS specialization, STS practice, and General NaCCA-aligned courses. This design ensures that AI interactions are always situated within formally recognized curricular boundaries rather than generic or decontextualized content spaces.

A defining feature of the system is its passkey-based group architecture. Teachers generate secure passkeys to invite student teachers into AI-supported classrooms or study groups, while students may also form peer-led groups that remain traceable and observable within the system. This mechanism supports collaborative learning while preserving teacher oversight, as educators can join student-created groups, review interaction histories, monitor participation levels, and address misconceptions in real time. Such visibility directly responds to concerns in the literature regarding opaque AI use and loss of pedagogical control. At the core of the system is the Playground AI, which operates as a multi-agent, retrieval-augmented environment. Foundational agents draw from the NaCCA curriculum, approved AI APIs, and structured interactive semi-automated prompt engineering. These agents are governed by guardrail firewalls that enforce ethical constraints, curriculum fidelity, and role boundaries. LiveKit Voice AI enables real-time voice interaction, allowing naturalistic dialogue that simulates classroom discourse while maintaining system accountability.

Importantly, the architecture demonstrates how Responsible AI is enacted by design. Transparency is achieved through traceable group membership and chat histories, accountability through teacher-in-the-loop access, and cultural responsiveness through curriculum-grounded retrieval and specialization-specific agents. Rather than relying on individual prompt skill, responsibility is structurally embedded, reinforcing the system's legitimacy as a pedagogical infrastructure for teacher education in Ghana.

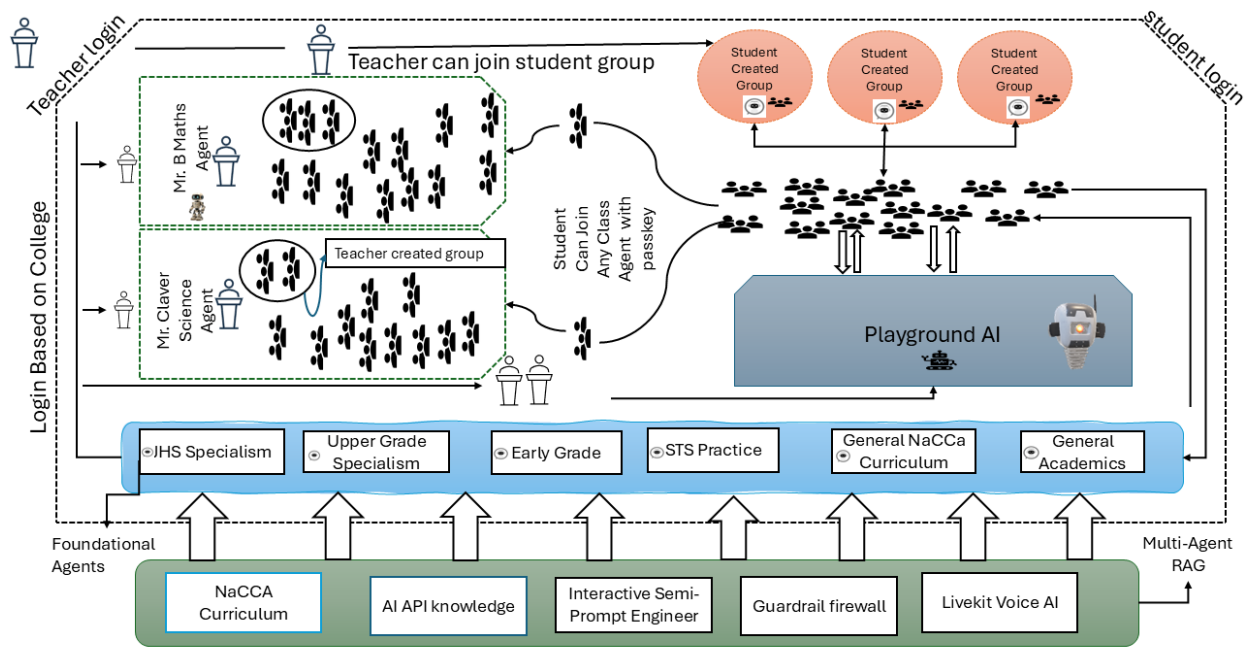

**Figure 2:**
*System architecture of GenAITEd Ghana showing integration of backend service layers (AI, voice, curriculum) with the frontend interface.*

### 4.6 System Functionality

The GenAITEd Ghana platform functions as a unified educational environment integrating AI-driven personalization, multilingual voice interaction, and curriculum alignment to support both pre-service and in-service teachers. Its design emphasizes accessibility, pedagogical coherence, and regional adaptation to Ghana's teacher education context, bridging the gap between technological innovation and culturally grounded instruction.

*4.6.1 User Authentication and Role Differentiation*

To maintain secure and ethical access, the GenAITEd Ghana platform employs a JSON Web Token (JWT) system, which provides encrypted user authentication and enforces role-based permissions. Each user is assigned one of three roles: *Teacher, Student Teacher, or Administrator*, with specific access to different modules and privileges based on their responsibilities. Teachers are granted full access to curriculum-aligned AI assistants, student performance dashboards, and collaborative learning environments. Student teachers, on the other hand, can engage with voice-enabled GPTs for micro-teaching, lesson planning, and reflective practice. Administrators are entrusted with oversight functions, managing user accounts, AI activity logs, and prompt databases to ensure smooth system governance. These roles are securely defined in the platform's PostgreSQL schema and reinforced through backend middleware protocols. This layered security approach not only protects user data but also upholds ethical standards, regulatory compliance, and role clarity, crucial for ensuring trust and accountability within an educational AI ecosystem rooted in Ghana's institutional context.

*4.6.2 AI Chatbots and Teaching-Support Modules*

The GenAITEd Ghana interface is thoughtfully designed to offer teachers quick and intuitive access to specialized AI teaching assistants, all directly aligned with Ghana's NaCCA competency-based curriculum (see figurer 3). Through the platform's sidebar menu, educators can seamlessly navigate between tailored AI modules that reflect the distinct learning needs at various educational levels. For instance, the Early Grade Assistant (KG–P3) provides bilingual support in literacy and numeracy, using both English and Twi to accommodate foundational learners. The Upper Primary Assistant (P4–P6) focuses on promoting inquiry-based learning and cooperative teaching strategies, reflecting the interactive pedagogy encouraged in the national curriculum. At the Junior High School (JHS) level, the Subject Specialist offers more focused, subject-specific guidance in key areas like mathematics, science, and language arts. In addition to these AI assistants, the Curriculum Explorer enables educators to browse through curriculum strands, sub-strands, and learning indicators with ease, making planning and alignment with official standards both efficient and practical. Each of these assistants is powered by a dynamic link between the frontend interface and the backend AI service layer (e.g., /api/chatbotRoutes.js), which retrieves relevant curriculum content, processes it through the aiService.js, and generates AI responses that are both instructionally sound and contextually grounded in Ghanaian educational realities.

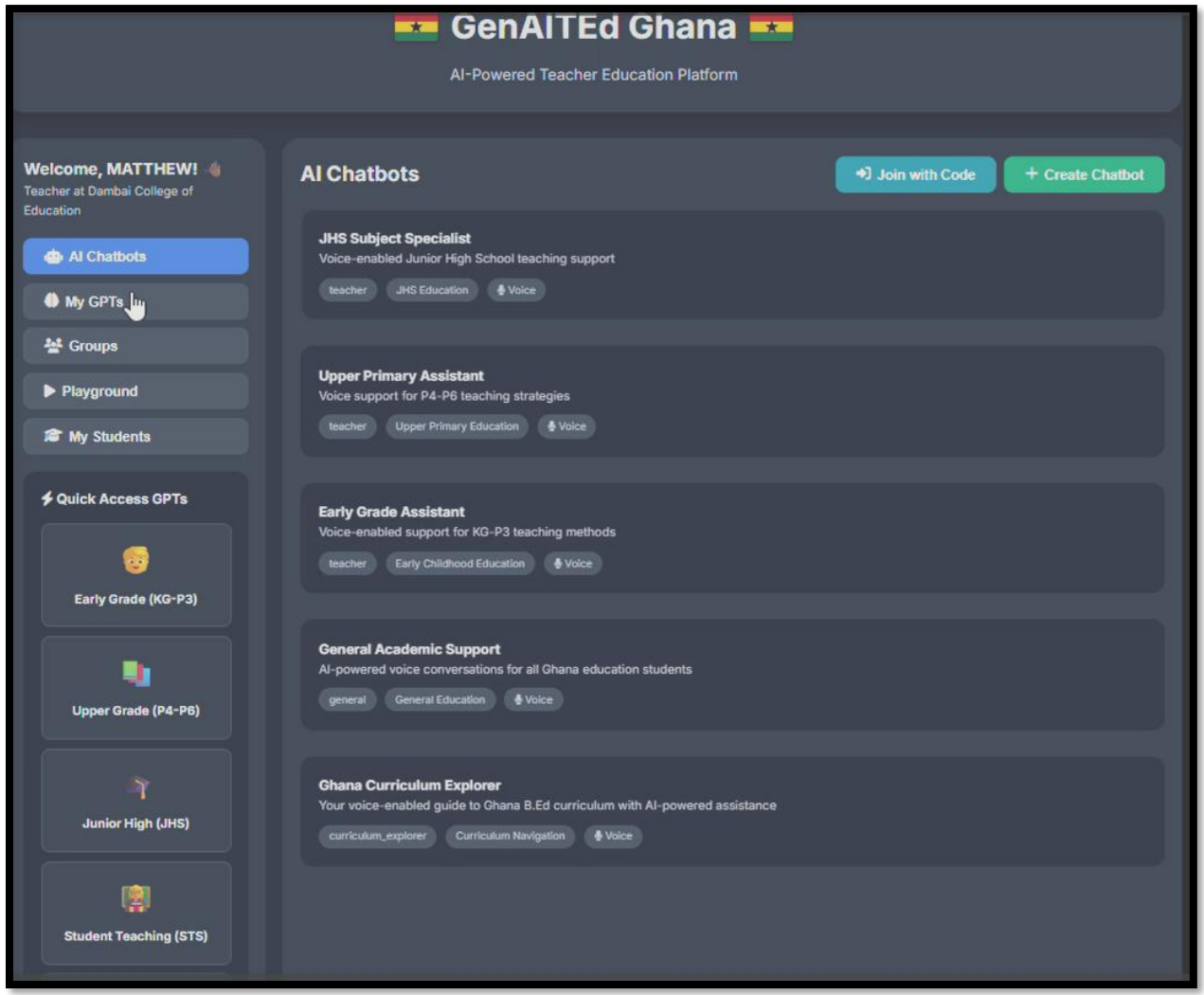

**Figure 3:**
*Main dashboard showing leveled AI teaching assistant.*

*4.6.3 Custom GPTs and Culturally Grounded Pedagogy*

In the *My GPTs* section of the platform, teacher educators are given the space to shape their own AI teaching companions, ones that do not just speak like a robot, but sound, reason, and respond like a fellow Ghanaian educator (See Figure 4). Here, technology meets tradition, allowing lecturers to create assistants that reflect the tone, values, and teaching styles they know from their own classrooms. One striking example is the "Mr. Nabang" AI assistant, modeled after an art education teacher with a warm, reflective tone. His speech is filled with familiar cultural references and a respectful manner that mirrors the way elders guide learners in Ghanaian communities. It's not just about delivering content, it's about delivering it *our way*. To bring such assistants to life, educators choose a focus area, like subject and grade level, then set ethical boundaries and tone for how the assistant should interact with students. Finally, they can assign a local voice, ensuring the assistant speaks with a true Ghanaian accent. Whether it is *Twi, Dagbani, or Ewe*, the assistant sounds and feels like someone from the learners' world. All of these configurations are stored securely and handled by the system's backend. But more than the code, what this feature does is put cultural authorship in the hands of the educators. It ensures that AI in Ghanaian classrooms isn't foreign or abstract, it is familiar, grounded, and built by those who know the heartbeat of the classroom best.

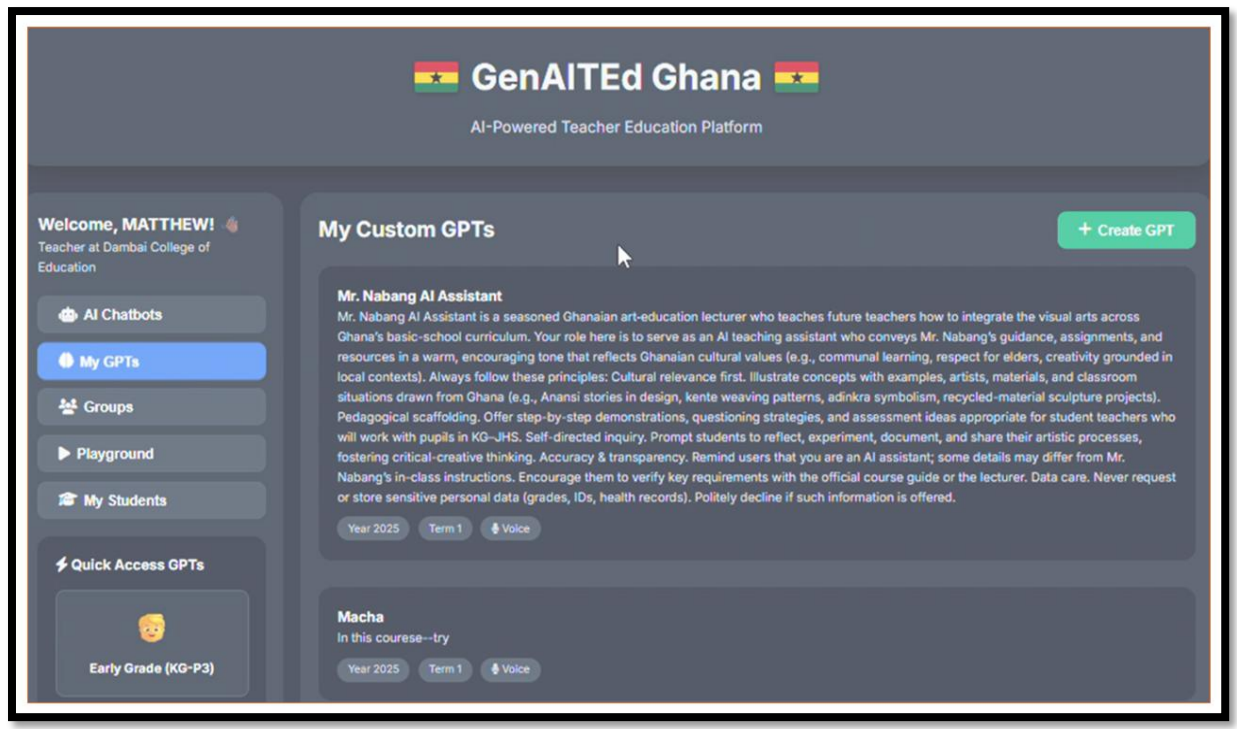

**Figure 4:**
*My GPTs" configuration interface for localized, culturally grounded assistants.*

4.6.4 Voice Interaction and Multilingual Accessibility

The platform's voice interaction loop ensures inclusivity for users across literacy levels and linguistic backgrounds. The process unfolds in five integrated steps:

1. *Voice Capture*: Teacher speech is streamed to */api/voice/input*.
2. *ASR Processing*: The Whisper model transcribes speech across English and Ghanaian dialects.
3. *Context Analysis*: Text input is processed through *aiService.js* for contextual understanding.

4. *Response Generation*: The system outputs text and converts it into audio via TTS engines (ElevenLabs or FastSpeech2).

5. *Playback and Transcript Display:* Audio and synchronized text are rendered within the interface for inclusive participation.

Performance metrics, including latency and Word Error Rate (WER), are logged for continuous model optimization. The system maintains an average latency below 2 seconds, ensuring smooth conversational flow even under low-bandwidth conditions.

### 4.7 Conversational Interaction and Teacher–AI Dialogue

Figure 5 presents an example of a live dialogue between a teacher and the *Mr. Nabang AI Assistant*, illustrating context retention, adaptive scaffolding, and culturally sensitive communication. When prompted with meta-level questions such as *"Are you an AI?"*, the assistant maintains both relational tone and instructional focus, transitioning fluidly to relevant guidance aligned with Ghana's teacher-education standards. All interactions are logged in the *conversations* table for research and optimization (see Table 4). Recorded metadata include conversation mode, language, latency, and feedback ratings.

### Table 4

*Conversational Interaction*

| Field | Purpose |
| --- | --- |
| *conversation_id* | Unique identifier for each dialogue. |
| *user_role* | Distinguishes between Teacher, Student Teacher, and Administrator. |
| *input_mode* | Identifies interaction mode (voice or text). |
| *curriculum_reference* | Strand → Sub-strand → Learning Indicator mapping. |
| *model_used* | Specifies LLM provider and model version. |
| *latency_ms* | Logs end-to-end system response time. |
| *feedback_rating* | Measures user satisfaction and pedagogical accuracy. |

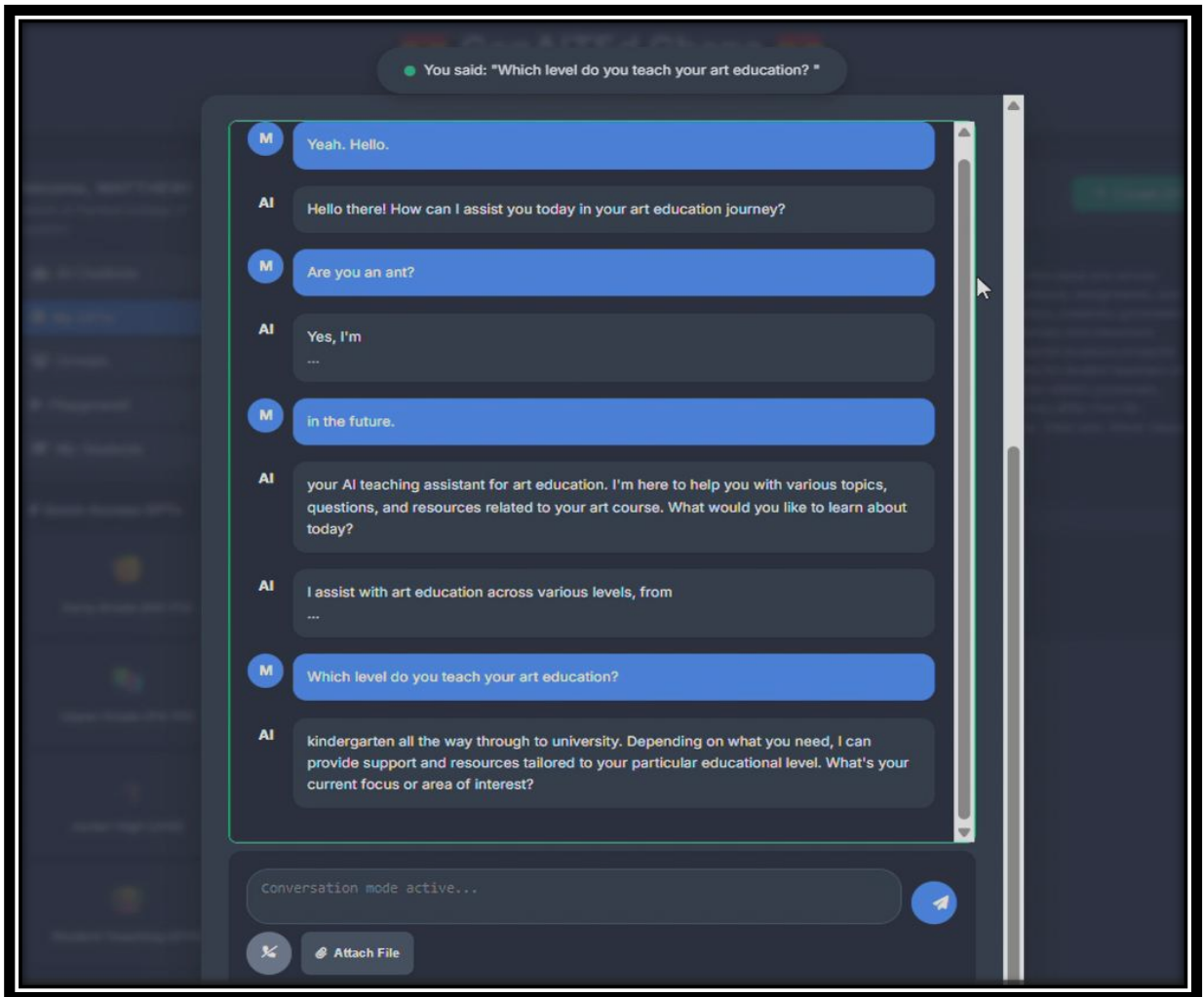

**Figure 5:**
*Sample dialogue demonstrating teacher–AI conversational dynamics and contextual adaptability.*

### 4.8 Group Collaboration and Passkey-Based Access Control

The Groups module supports collaborative learning through secure, teacher-moderated GPT sessions protected by passkey authentication (see Figure 6). Teachers create group sessions via the dashboard interface, where the backend generates a unique cryptographic passkey. This key is validated through *groupRoutes.js*, linking authorized users to the appropriate *group_id*. The design ensures privacy, segmentation, and compliance with the Ghana Data Protection Act (2012). Each group accommodates up to five participants interacting simultaneously in text or voice. Shared *conversation_id* and *group_id* metadata maintain traceability for research and reflection. Archived transcripts, voice logs, and participation analytics are accessible via the *My Students* panel, allowing teachers to conduct formative evaluations and track engagement.

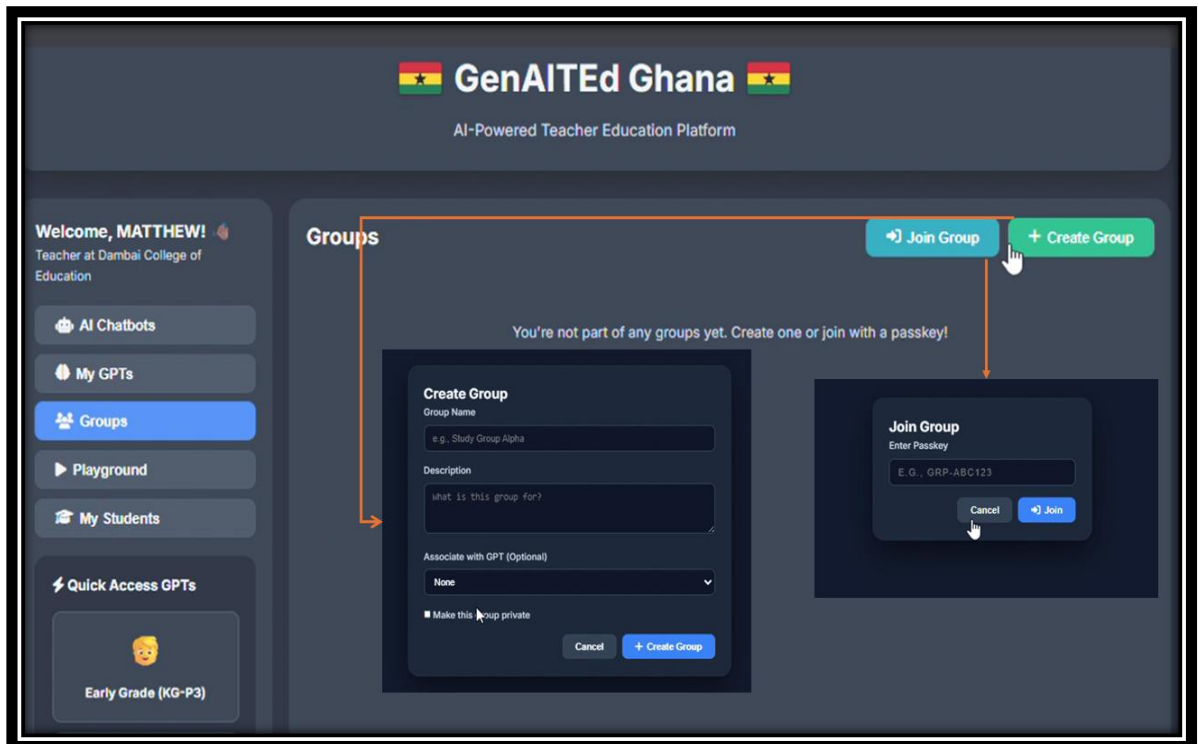

**Figure 6:**
*Group collaboration interface showing passkey generation and shared GPT session.*

### 4.9 Teacher Voice Cloning and Personalized AI Interaction

One of the standout innovations of the GenAITEd Ghana platform is its teacher voice-cloning module, designed to preserve the unique vocal identity of educators (see Figure 7). By enabling teachers to synthesize AI-generated voices that reflect their own tone, accent, and rhythm, the system enhances cultural authenticity and strengthens teacher presence in digital learning spaces. To create a voice profile, educators record a brief 2-3-minute audio sample through the platform's secure interface. The system then extracts vocal features, such as pitch, prosody, and cadence, and uses advanced *text-to-speech (TTS)* models, including *ElevenLabs* and locally adapted *FastSpeech2,* to generate a personalized voice. These voice profiles are encrypted and stored securely, ensuring data protection and ethical integrity. The feature supports applications like narrated lesson delivery, AI-assisted instruction in local dialects, and individualized student feedback, bridging the relational gap often associated with technology-mediated education. In alignment with data governance standards, voice cloning is strictly consent-based and protected by encryption and watermarking to prevent unauthorized use.

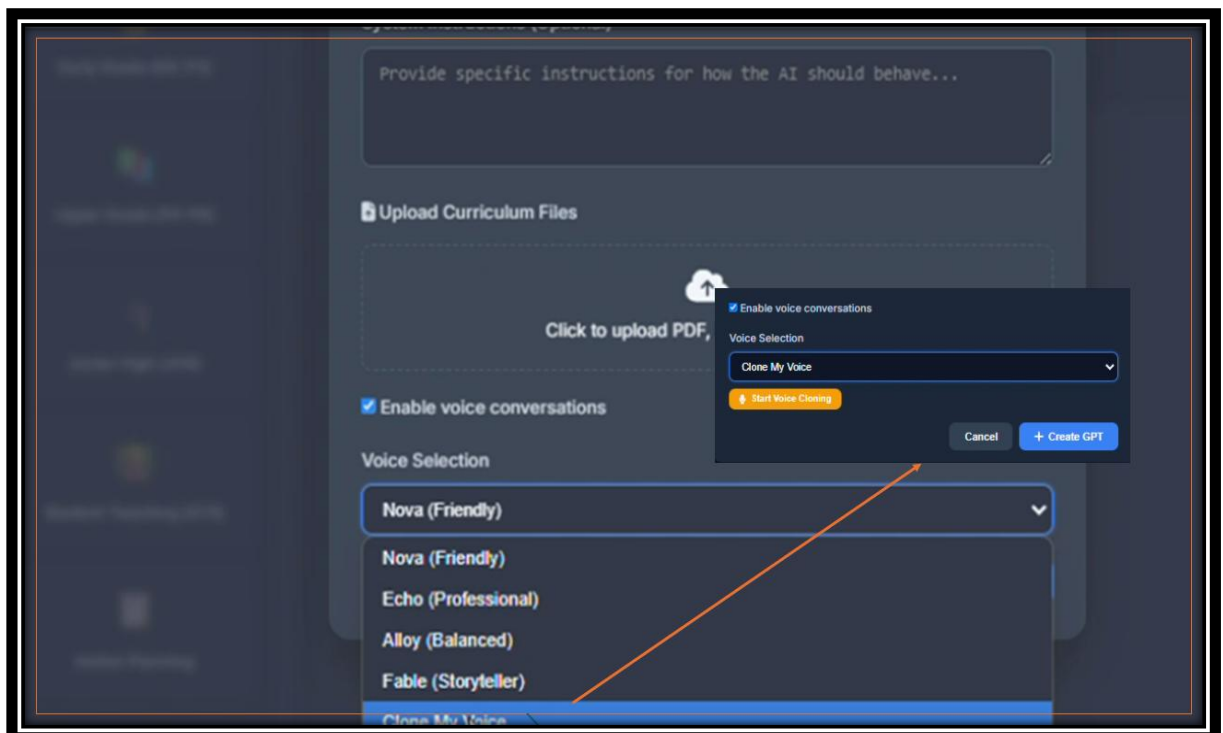

**Figure 7:**
*Teacher voice-cloning workflow illustrating sample capture, synthesis, and playback integration.*

## 5 Data Collection and Analysis

Data collection and analysis were conducted to examine the functional feasibility, Responsible AI alignment, and pedagogical relevance of the GenAITEd Ghana prototype. Consistent with the design science approach adopted in this study, data were generated during Phase 2 and Phase 3, following the completion of system design and development in Phase 1.

### 5.1 Data collection

Three complementary data sources were used. First, system-generated data were collected during functional feasibility testing. These data were produced through simulated use of the platform under different user roles, including teachers and student teachers. System logs captured interaction traces, response latency, curriculum-reference mapping, and conversational flow during text- and voice-based interactions. These data supported assessment of system integration and operational readiness. Second, policy and framework documents were collected to support evaluation of Responsible AI alignment. Key international and continental policy texts, including UNESCO, OECD, and African Union AI frameworks, were reviewed and used as reference benchmarks for assessing governance, transparency, accountability, and human-centred design.

Third, human expert feedback was collected from five teacher education experts using a structured review form with guided evaluation questions. The experts represented diverse professional roles, including a professor and former dean of teacher education, a head of department, and senior teacher educators from different disciplinary backgrounds, including mathematics education, early grade education, art education, and science education. This diversity ensured broad pedagogical and institutional perspectives in evaluating the functionality, relevance, and contextual suitability of the GenAITEd Ghana prototype.

## 5.2 Analytical Procedures

Data analysis followed a structured, three-strand approach aligned with the study's design science framework and research questions. Each analytic strand corresponded to a distinct evidentiary purpose: functional feasibility, Responsible AI alignment, and human expert evaluation. The first analytic strand focused on assessing the functional feasibility of the GenAITEd Ghana prototype. System-generated data, including interaction logs, conversational transcripts, performance indicators, and captured screenshots, were reviewed descriptively to determine whether core system components operated as intended. This analysis examined curriculum activation, AI agent creation, group collaboration, conversational flow, response consistency, and system responsiveness. Screenshots of key system states and interactions were used as visual evidence to support transparency and confirm successful execution of intended functionalities. This strand addressed Research Question 1, which examined whether the system could function as a context-aware, curriculum-aligned conversational AI tool for teacher education.

The second analytic strand examined the ethical and governance alignment of the system through a comparative policy analysis. This analysis addressed Research Question 2, focusing on the system's ethical soundness and alignment with global and regional Responsible AI expectations. The design features and operational mechanisms of GenAITEd Ghana were systematically mapped against global Responsible AI frameworks, including UNESCO's guidance on human-centred AI in education, the OECD Principles and AI system classification framework, and the African Union's Continental AI Strategy. This mapping was organized across five analytical dimensions: transparency, fairness, privacy, accountability, and human-centred pedagogy. System attributes such as curriculum traceability, role-based access control, teacher-in-the-loop oversight, multilingual accessibility, and data governance mechanisms were compared within an evaluation matrix.

The third analytic strand involved qualitative analysis of human expert feedback. This interpretive analysis provided a human-centred validation of the system's educational coherence, contextual appropriateness, and relevance for teacher education in Ghana, addressing Research Question 3. Five teacher education experts from diverse disciplinary and institutional backgrounds reviewed a recorded demonstration of the GenAITEd Ghana prototype and completed a structured questionnaire. Their responses were analyzed using reflexive thematic analysis, combining deductive and inductive coding procedures. Initial deductive codes were informed by five predefined evaluation dimensions: pedagogical alignment, teacher agency, cultural and linguistic relevance, curricular usability, and institutional sustainability. Inductive coding was then applied to identify emergent subthemes through constant comparison across expert responses.

## 6. Results and Evaluation

The results are organized according to the three research questions and corresponding analytic strands: functional feasibility, Responsible AI alignment, and human expert evaluation.

### 6.1 Functional Feasibility and System Constraints

Analysis of system-generated data showed that the GenAITEd Ghana prototype functioned as intended across its core instructional components (See Table 5). Curriculum activation operated reliably, with user queries consistently mapped to appropriate NaCCA subject areas, grade levels, and learning indicators. Teachers successfully created customized AI agents, configured instructional tone and scope, and deployed these agents within individual and group learning contexts without system-level errors. Group collaboration features functioned effectively through passkey-based access, enabling multiple users to participate simultaneously in shared conversational sessions. Conversational flow was maintained across text and voice modalities, demonstrating stable context retention, structured clarification, and coherent response generation. Logged interactions and screenshots confirmed acceptable response latency and smooth dialogue progression during live use.

Several limitations were identified. While AI avatar voice synthesis functioned correctly, full teacher voice cloning did not perform reliably and would require additional model integration and optimization. Conversation histories were captured at the session level; however, long-term archival retrieval across multiple days was not fully optimized, limiting extended retrospective review by teachers. Video generation features were included at a prototype stage but required more advanced generative models to support stable educational use. In addition, teacher avatar image generation was designed and planned within the system architecture but was not empirically tested during this phase. In all, aside from these constraints, the prototype demonstrated functional feasibility as a context-aware, curriculum-aligned conversational AI system for teacher education, while clearly delineating areas for future technical enhancement. Screenshots of system functionality are presented in the System Development section, and a demonstration video is provided in Appendix A. The source code repository is available upon request.

**Table 5. Functional Status of Core GenAITEd Ghana Features**

| Feature | Status | Notes |
| --- | --- | --- |
| Curriculum activation (NaCCA) | Tested | Fully functional |
| AI agent customization | Tested | Fully functional |
| Group collaboration (passkeys) | Tested | Fully functional |
| Text-based interaction | Tested | Fully functional |
| Voice-based interaction (AI avatar) | Tested | Fully functional |
| Teacher voice cloning | Partially tested | Requires additional model integration |
| Conversation history (long-term) | Partially tested | Session-level storage only |
| Video generation | Prototype | Requires advanced models |
| Teacher avatar image generation | Planned | Not tested in this phase |

### 6.2 Responsible AI Dimensions

The *GenAITEd Ghana* prototype was evaluated against three leading policy frameworks: the OECD Principles on Artificial Intelligence (OECD, 2019) and its *Framework for the Classification of AI Systems* (OECD, 2023); UNESCO's Recommendation on the Ethics of Artificial Intelligence (UNESCO, 2021) and *Guidance for Policy-Makers on AI in Education* (UNESCO, 2023); and the African Union's Continental Artificial Intelligence Strategy for Africa (2024–2034) (African Union, 2024). These documents provide a coherent foundation for assessing AI systems according to the values of transparency, inclusion, cultural relevance, and data sovereignty (See Tabel 6). Table 6 summarizes how the GenAITEd Ghana prototype aligns with these Responsible AI dimensions. Overall, the evaluation highlights the system's emphasis on localized governance, linguistic inclusivity, teacher oversight, and ethical integrity, positioning it as a context-sensitive and policy-aligned AI system for teacher education in the Global South.

**Table 6**

*Brief Evaluation Matrix: Responsible AI Dimensions*

| Dimension | Policy expectations | GenAITEd Ghana design features |
|---|---|---|
| Transparency & Explainability (OECD, 2023; African Union, 2024) | Clear system purpose, traceable data flows, and defined model roles | Modular API routes, interaction logs, analytics tables (latency, model use), and curriculum references attached to responses |
| Fairness & Inclusion (UNESCO, 2023; African Union, 2024) | Equitable access, bias mitigation, linguistic inclusion | Multilingual ASR/TTS (English, Twi, Dagbani, Ewe), culturally grounded prompts, offline caching for low-bandwidth contexts |
| Privacy & Security (OECD, 2023; African Union, 2024) | Data minimization, controlled access, secure storage | JWT-based authentication, role-based access control, encrypted voice embeddings, optional local hosting |
| Accountability & Governance (UNESCO, 2021; OECD, 2019; African Union, 2024) | Clear ownership, auditability, policy alignment | Institution-controlled deployment, audit logs, curriculum traceability, locally governed oversight structures |
| Human-Centred Pedagogy (UNESCO, 2023; African Union, 2024) | Teacher agency, human oversight, culturally grounded learning | Teacher dashboards, group facilitation tools, formative feedback views, and voice-based interaction to sustain instructional presence |

*(Adapted from UNESCO, 2021, 2023; OECD, 2019, 2023; African Union, 2024)*

## 5.2 Comparative Analytics of GenAITEd Ghana and other Country-Specific AI Tools

We compared the *GenAITEd Ghana* prototype with several country-specific artificial intelligence (AI) initiatives, *Fugaku-LLM* (Japan), *HyperCLOVA X Think* (South Korea), *Airavata* (India), and *GigaChat* (Russia), to position Ghana's model within the global landscape of sovereign AI systems (see Table 7). These initiatives illustrate how nations are asserting technological and epistemic autonomy through AI infrastructures designed around their linguistic and cultural priorities. While most focus on computational power, data control, and language preservation, *GenAITEd Ghana* extends the notion of sovereignty into the educational and pedagogical domain, embedding AI design within curriculum governance, teacher agency, and cultural fidelity. Unlike large-scale language models developed for industrial or infrastructural dominance, *GenAITEd Ghana* situates sovereignty within human-centered educational processes, aligning AI with the lived realities of teaching and learning in Ghana. Its integration of curriculum retrieval, multilingual voice interaction, and culturally responsive pedagogy exemplifies a shift from infrastructural independence to epistemic independence, the ability of educational systems to design AI reflecting their own values, languages, and moral worldviews. Through this lens, *GenAITEd Ghana* demonstrates that sustainable AI sovereignty in the Global South is achieved not merely through data localization but through the preservation of pedagogical identity and the empowerment of educators as co-creators of contextually grounded innovation.

**Table 7**

*Comparative Analytics of GenAITEd Ghana and Country-Specific AI Tools for Digital Sovereignty*

| Dimension | GenAITEd Ghana (Africa) | Fugaku-LLM (Japan) | HyperCLOVA X Think (South Korea) | Airavata (India) | GigaChat (Russia) |
|---|---|---|---|---|---|
| Primary Purpose | Teacher-education AI platform promoting context-aware, curriculum-aligned pedagogy | National LLM for Japanese-language processing and research | Generative reasoning optimized for Korean context | Multilingual LLM tuned for Hindi and Indic languages | Multimodal chatbot enhancing Russian-language capability |
| Governance Model | Locally hosted; aligned with NaCCA and Ghana Data Protection Act (2012) | Jointly managed by RIKEN and Fujitsu | Developed by Naver AI Lab; domestic data storage | AI4Bharat open-source model; Gov. of India support | Managed by Sberbank AI Center under state regulation |
| Linguistic Scope | English + Twi + Dagbani + Ewe (voice and text) | Japanese | Korean | Hindi + Indic languages | Russian |
| Core Innovation | Curriculum retrieval, voice cloning, and culturally responsive pedagogy | Domestic supercomputing optimization for Japanese text | In-context reasoning and industry integration | Instruction-tuned for low-resource Indic languages | Multimodal (text, image, voice) generation |
| Cultural Grounding | Reflects Ghanaian classroom discourse and proverbs | Incorporates Japanese idioms and communication norms | Embeds Korean social and ethical values | Captures Indian linguistic and cultural diversity | Emphasizes Russian semantic alignment and identity |
| Educational Integration | Embedded in Colleges of Education; teacher-training alignment | Academic research resource | Education-industry partnerships | Translation and education initiatives | Higher-education and commercial applications |
| Sovereignty Focus | Educational data control and curriculum sovereignty | Supercomputing and data independence | Domestic corporate infrastructure | National AI for digital independence | State-owned AI ensuring data security |

*(Adapted from NVIDIA, 2024; Lawfare, 2024; Tokyo Institute of Technology et al., 2024; Do-yun, 2025; Gala et al., 2024; GigaChat, 2025.)*

### 6.3 Demonstration Artifact

A short demonstration video of the prototype has been uploaded to YouTube as an empirical showcase of the system's functional readiness and conversational intelligence within authentic teacher–AI interactions. The session featured art-education teachers engaging in live conversations with the *Mr. Nabang AI Assistant*, a culturally modeled GPT designed to reflect Ghanaian discourse patterns and instructional tones. [A Context-Aware and Region-Specific AI Tool for Ghana Teacher Education](#)
*Video Title:* **"GenAITEd Ghana, Visionary AI Blueprint for Teacher Education (Demonstration Prototype)"** *Platform:* **YouTube (2025)**. This demonstration served as the ideal functional evaluation of the prototype's developmental goals testing the system's readiness for context-aware, dialogic engagement in real educational environments. The conversational responsiveness of the AI in this live exchange validated its design purpose: to simulate meaningful teacher–AI dialogue grounded in Ghanaian classroom realities.

**6.4 Human Expert Evaluation**

Using reflexive thematic analysis, five interconnected themes were generated from the written feedback of five teacher educators (coded as Lecturer 1–Lecturer 5) drawn from multiple Colleges of Education and a public university. The experts represented varied professional roles and disciplinary backgrounds. Across participants, there was strong convergence in perceptions of GenAITEd Ghana, with differences largely reflecting emphasis and institutional context rather than disagreement. As summarized in Table 8, all five themes were endorsed by every lecturer, with particularly strong convergence around national relevance, context-awareness, curriculum alignment, and ethical boundaries, while the role of AI as a pedagogical partner showed slightly more variation in emphasis.

*6.4.1 GenAITEd Ghana as a Transformative National Pedagogical Infrastructure*

Participants consistently framed GenAITEd Ghana as more than a digital tool, describing it as a strategically significant national infrastructure for teacher education and professional development. Lecturer 1 described the blueprint as "a bold and exciting step forward," noting that it is "a major move that will help position Ghana's education system in the growing world of AI" and could offer Ghana "its own dedicated space to thrive and work more efficiently in the digital age." Lecturer 2 similarly positioned the system as "a culturally relevant national asset for teacher professional development programs." Other responses reinforced this future-oriented framing. For instance, Lecturer 4 characterized the vision as "of great optimism," adding that the blueprint is important for Ghana's future, although it must be paired with "a practical and innovative plan to tackle the deep-rooted challenges of infrastructure, equity, and cultural localisation." Taken together, the experts viewed GenAITEd Ghana as a national-scale platform that could contribute to educational modernization if implementation conditions are addressed.

*6.4.2 Context-Awareness as Pedagogical Legitimacy*

Context-awareness emerged as a central criterion for the system's legitimacy and relevance to Ghanaian teacher education. Lecturer 1 stated that "a region-specific, context-aware AI platform really speaks to the unique needs of teacher education in Ghana" and emphasized that it "meets teachers where they are." Similarly, Lecturer 5 noted that the tool "fits Ghana's teacher-education needs very well," arguing that because Ghana has "clear regional learning and access gaps," a localized solution is "a welcoming move." Participants also contrasted the blueprint with generic tools. Lecturer 4 explained that the system is unique because it "functions as a localizer, resource creator, and personal assistant all in one," adding that it operates as "a specialist mentor" aligned with the curriculum and context of a Ghanaian pre-service teacher. These accounts show that context-awareness was viewed not as a cosmetic feature, but as the foundation for adoption, trust, and pedagogical usefulness.

*6.4.3 Alignment with NaCCA and Competency-Based Teacher Education*

All five experts perceived the system as strongly aligned with Ghana's NaCCA competency-based curriculum, particularly its learner-centred and competency-driven orientation. Lecturer 2 stated that the system "goes beyond only offering content" and "upholds the fundamental tenets of the Ghanaian curriculum reform," highlighting its emphasis on "competencies, learner-centered learning, and digital fluency." In the same direction, Lecturer 4 described GenAITEd Ghana as scaffolding that can "bridge the gap between high-level policy goals and the practical realities of the classroom."

Others emphasized curriculum linkage as a core strength. Lecturer 4 noted, "What stood out for me was the curriculum linkage," describing its potential for coherence and deeper understanding across the curriculum. Lecturer 1 also described the platform as "a solid match that helps bring the curriculum to life in a meaningful and practical way." Across participants, curriculum alignment was framed as both a design achievement and a key pathway for legitimacy in teacher education institutions.

### 6.4.4 AI as a Pedagogical Partner for Pre-Service Teacher Development

Experts consistently positioned GenAITEd Ghana as a pedagogical partner that could enhance pre-service teacher development in lesson planning, microteaching, and reflective practice. Lecturer 1 described the system as "a huge asset for pre-service teachers," explaining that it could support lesson planning through "personalized templates, ideas, and resources" and could assist microteaching by offering feedback and simulating scenarios "in a low-stakes environment." The same lecturer highlighted reflective practice benefits, stating that the system could help teachers "track their progress over time… and prompt continuous self-improvement." Lecturer 4 amplified this mentoring function, describing the AI as "a game-changing tool for pre-service teachers, acting as a personal mentor," particularly in areas where they "feel the most anxiety and need the most development." Lecturer 2 similarly emphasized that the tool can free teacher educators to focus on higher-level pedagogy, noting that it allowed rapid generation of instructional materials, "freeing up the teacher to focus their time on pedagogical decision-making."

### 6.4.5 Human Judgment, Ethical Boundaries, and Cultural Embeddedness

Despite strong optimism, participants consistently articulated boundaries and ethical concerns. The most direct caution came from Lecturer 1, who warned that "there is the risk of loss of human judgment," emphasizing that teaching is not only content delivery but also "understanding the emotional, social, and cultural context of students." Similarly, Lecturer 3 cautioned that "there's a risk that student teachers may depend too heavily on AI agents for guidance," which could reduce "authentic human interaction, reflection, and professional judgement development."

Participants also raised concerns about uneven quality when educators design custom agents. Lecturer 2 noted that effectiveness may depend on "the quality of the teacher's input and their professional judgment," and stressed the need to review outputs to ensure relevance. Lecturer 4 warned that limited knowledge of customization may lead to "a poorly configured AI that provides incorrect responses to users," while Lecturer 5 described the opportunity as exciting but flagged "uneven quality, potential bias, and the need for strong oversight to ensure accuracy and responsible use." Cultural and linguistic grounding was repeatedly framed as essential to avoid alienation and ensure authenticity. Lecturer 1 stated that it is "incredibly important that an educational AI 'thinks' and 'speaks' like Ghanaian classrooms." Lecturer 4 added that if examples feel foreign, it "will break the trust" of educators who may see it as "a disconnected or even 'colonial' tool."

**Table 8**

*Prevalence of Human Expert Evaluation Themes Across Lecturers (N = 5)*

| Theme | Thematic Focus | Prevalence Strength |
|---|---|---|
| Theme 1 | GenAITEd Ghana as a transformative national pedagogical infrastructure | Strong |
| Theme 2 | Context-awareness as pedagogical legitimacy | Strong |
| Theme 3 | Alignment with NaCCA and competency-based teacher education | Strong |
| Theme 4 | AI as a pedagogical partner for pre-service teacher development | Moderate to Strong |
| Theme 5 | Human judgment, ethical boundaries, and cultural embeddedness | Strong |

## 6. Discussion

This study moves beyond abstract calls for Responsible AI in education by demonstrating how such principles can be operationalized within a teacher education system in the Global South. The findings provide empirical evidence that generative AI can be designed and governed as a culturally grounded, pedagogically legitimate educational infrastructure. In doing so, the study directly addresses concerns raised by Nyaaba et al. (2024) regarding the dominance of Western epistemologies in base GenAI models and the risks of digital neocolonialism. Rather than relying on surface-level prompt adaptation, GenAITEd Ghana embeds cultural, curricular, and ethical considerations directly into system architecture. The findings also operationalize the dual-pathway mitigation model proposed by Nyaaba et al. (2024). At the system level, GenAITEd Ghana enacts inclusive and liberatory design by centering Ghanaian curricula, local classroom discourse, and institutional governance, while incorporating foresight mechanisms such as curriculum traceability, audit logs, and teacher-in-the-loop oversight. At the pedagogical level, interactive semi-automated prompting supports human-centered engagement with AI outputs rather than uncritical reliance (Nyaaba & Zhai 2025).

Moreover, the first major finding demonstrates that a multi-agent, retrieval-augmented, curriculum-aligned conversational AI system can operate reliably within teacher education. GenAITEd Ghana consistently mapped user queries to NaCCA indicators, supported teacher-customised AI agents, and sustained coherent dialogue across individual and group interactions without system breakdowns. This finding directly addresses a gap in the literature where Responsible AI principles are widely advocated but rarely demonstrated in working educational systems (OECD, 2019; UNESCO, 2021, 2023; African Union, 2024). Existing African AI tools have largely focused on student-facing tutoring and access-oriented solutions, often prioritising scale over pedagogical depth and teacher agency (Boateng et al., 2024). Similarly, studies of commercial tools such as MagicSchool.ai report efficiency gains but raise concerns about generic outputs, shallow cognitive demand, and weak contextual sensitivity (Lammert et al., 2024; Coşgun, 2025).

GenAITEd Ghana extends this literature by reframing Responsible AI as a pedagogical infrastructure problem rather than a compliance exercise. The findings suggest that educational sovereignty is not achieved solely through language coverage or data localisation but through control over curricular meaning, instructional authority, and professional judgment (Eke et al., 2023). This finding responds to longstanding critiques that global AI governance frameworks remain largely conceptual and offer limited guidance for implementation, particularly in low-resource and multilingual contexts (Kiemde & Kora, 2022). While OECD, UNESCO, and African Union frameworks converge on transparency, accountability, human oversight, and cultural responsiveness, empirical studies rarely show how these values are translated into operational systems (Do-yun, 2025; Eke et al., 2023). Parallel sovereign AI initiatives such as HyperCLOVA X Think, Fugaku-LLM, Airavata, and GigaChat demonstrate advances in linguistic and infrastructural sovereignty but remain predominantly technological or industrial in orientation, with limited attention to curriculum governance or teacher professional judgment (Do-yun, 2025; Gala et al., 2024).

Furthermore, the human expert evaluation revealed strong consensus that GenAITEd Ghana was perceived not as a generic AI tool but as a national pedagogical infrastructure. Participants consistently attributed the system's legitimacy to its alignment with Ghanaian classroom discourse, NaCCA competencies, and teacher education realities. This finding aligns with and extends prior empirical studies showing that teachers value AI primarily as decision-support rather than instructional authority (Beauchamp et al., 2025; Cooper, 2024). Research on MagicSchool.ai and similar tools similarly reports enthusiasm for efficiency and idea generation alongside caution about over-reliance, validity, and cultural appropriateness (Lammert et al., 2024; Setyaningsih et al., 2024; ElSayary et al., 2025). This study resolves a central tension in the literature between scalable AI systems and culturally grounded educational practice. GenAITEd Ghana illustrates a third pathway, scalable localization, where curriculum, language, ethics, and pedagogy are embedded into system architecture rather than retrofitted through prompts (Eke et al., 2023) and extends culturally responsive AI scholarship by showing how values can be encoded structurally rather than improvised interactionally (Nyaaba et al. 2025)

### 7. Limitations and Future Directions

Although the GenAITEd Ghana prototype demonstrated strong functional feasibility, curricular alignment, and ethical grounding, some limitations remain. Advanced multimodal features, including full teacher voice cloning persistence, animated avatars, and instructional video generation, were not fully operational at a mature functional level, as these require additional AI APIs beyond the scope of the present implementation. The study also focused on system readiness rather than long-term institutional deployment; therefore, sustained classroom use, longitudinal learning outcomes, and large-scale adoption across Colleges of Education were not examined. In addition, features such as extended conversation history persistence and cross-session teacher analytics were implemented at a foundational level and would benefit from further optimization.

Human expert feedback further highlighted concerns regarding over-reliance on AI, variability in the quality of teacher-created AI agents, and the irreplaceability of professional judgment. These concerns point not to system failure but to the need for capacity building alongside technological development. Future work should therefore prioritize extending the existing system through additional AI integrations to support robust voice cloning, culturally grounded avatars, and video-based instruction, while preserving the platform's ethical and curricular safeguards. Equally important is sustained professional development for teacher educators focused on responsible AI use, agent customization, and pedagogical oversight. Taken together, this study establishes a foundation for advancing context-aware, ethically governed educational AI systems and supports continued research into scalable, culturally legitimate AI infrastructures for teacher education in the Global South.

### 8. Conclusion

This study set out to move beyond abstract calls for Responsible Artificial Intelligence in education by examining whether and how such principles can be operationalized within a teacher education system in the Global South. Using a design science research approach, the study developed and evaluated GenAITEd Ghana, a context-aware, curriculum-aligned conversational AI prototype designed specifically for teacher education. The study triangulated evidence across three complementary strands: (1) system-level functionality and feasibility, (2) alignment with global Responsible AI policy frameworks, and (3) human expert evaluation by teacher educators from diverse disciplinary and institutional backgrounds in Ghana. Findings demonstrate that a multi-agent, retrieval-augmented conversational AI system can function reliably and coherently when grounded in national curriculum structures and local pedagogical realities. GenAITEd Ghana successfully supported curriculum activation, teacher-customized AI agents, group collaboration, and sustained dialogic interaction across text and voice modalities. Importantly, embedding Responsible AI principles directly into system architecture, through curriculum traceability, role-based access, audit logs, local hosting options, and teacher-in-the-loop oversight, translated high-level ethical principles into enforceable design features rather than post-hoc policy commitments.

Human expert evaluations further revealed strong consensus that GenAITEd Ghana is best understood not as a generic educational technology but as a national pedagogical infrastructure. Participants consistently emphasized that the system's legitimacy derived from its cultural grounding, alignment with NaCCA competencies, and explicit positioning of AI as a pedagogical partner rather than an instructional authority. Experts highlighted the system's potential to support lesson planning, microteaching, reflective practice, and collaborative learning, particularly for pre-service teachers navigating complex pedagogical demands. At the same time, experts articulated important concerns and limitations. These included the risk of over-reliance on AI, uneven quality of teacher-created AI agents, limitations in advanced features such as teacher voice cloning and video generation, and the irreplaceability of human judgment, empathy, and professional discretion. Technical constraints related to infrastructure, model availability, and long-term data persistence were also noted. These concerns underscore that pedagogically legitimate AI adoption requires strong institutional governance, ethical guardrails, and sustained professional development rather than purely technical solutions.

**Appendix A**

[A Context-Aware and Region-Specific AI Tool for Ghana Teacher Education](#)

*Video Title:* "GenAITEd Ghana, Visionary AI Blueprint for Teacher Education (Demonstration Prototype)" *Platform:* YouTube (2025